\documentclass[12pt]{article}
\usepackage{graphics}
\def\ea{{\em et al.}}
\def\ni{\noindent}
\def\ph{{\phantom{...}}}
\def\={\phantom{..} = \phantom{..}}

\def\+{\phantom{..} + \phantom{..}}
\def\>{\phantom{..} > \phantom{..}}
\def\<{\phantom{..} < \phantom{..}}
\def\-{\phantom{..} - \phantom{..}}

\def\lra{\longrightarrow}

\def\rng{random number generator}
\def\inter{intervention}
\def\ie{intervention-efficacy}

\def\pop{population}
\def\pops{populations}

\def\rv{random variable}

\def\bp{branching-process}
\def\bps{branching-processes}

\def\prob{probability}
\def\probs{probabilities}

\def\epv{extra-Poisson variation}

\def\ro{R_0}
\def\vo{V_0}
\def\rl{R_l}
\def\rh{R_h}
\def\ph{p_h}
\def\pl{p_l}
\def\pe{p_e}
\def\pei{p_{e,i}}

\def\no{\nonumber}
\def\be{\begin{equation}}
\def\ee{\end{equation}}

\def\bar{\begin{eqnarray}}
\def\ear{\end{eqnarray}}
\def\bad{\begin{equation} \begin{array}}
\def\ead{\end{array} \end{equation}}
\def\bmat{\left( \begin{array}}
\def\emat{\end{array} \right)}

\title{\bf \scalebox{1.5}{Stopping the SuperSpreader}\\ 
\scalebox{1.5}{Epidemic:}\\[1in] 
the lessons from SARS (with, perhaps, applications to MERS)\\[2in]}

\author{W. David Wick$^1$\footnote{
Email:wdavid.wick@gmail.com}} 

\begin{document}
\maketitle
\pagebreak

\section*{Abstract}
I discuss the so-called SuperSpreader (SS) epidemic, for which SARS is the canonical example (and, perhaps, MERS will be another).
I use simulation by an agent-based model as well as the mathematics of multi-type \bps\ to illustrate how 
the SS epidemic differs from the more-familiar uniform epidemic (e.g., caused by influenza).
The conclusions may surprise the reader: (a) the SS epidemic must be 
described by {\em at least two numbers}, such as the mean reproductive number (of ``secondary'' infections caused by a ``primary'' case), $\ro$, {\em and} the variance of same,
call it $\vo$; (b) Even with $\ro > 1$, if $\vo \gg \ro$ the probability that the infection-chain caused 
by one primary case goes extinct without intervention may be close to one (e.g., 0.97); (c) The SS epidemic
may have a long ``kindling period'' in which sporadic cases appear (transmitted from some unknown host) and generate a cluster of cases, 
but the chains peter out, perhaps generating 
a false sense of security that a pandemic will not occur; (d) Interventions such as isolating primary cases (or contact-tracing and secondary-case isolation) can be efficacious
even without driving $\ro$ below one; (e) The efficacy of such interventions diminishes, but slowly, with increasing $\vo$ at fixed $\ro$. From these considerations, I argue that the
SS epidemic has dynamics sufficiently distinct from the uniform case that efficacious public-health interventions can be designed even in the absence of a vaccine or other
form of treatment. 
\pagebreak

\section*{Text}
This is a mathematical-modeling paper about certain types of epidemics and corresponding public-health interventions, other than by vaccines or drugs. 
My desire is to communicate the main points to a wider audience then just my fellow
math-modelers: in particular, infectious-disease doctors and public-health officials, who are charged with responding to, and, if possible, halting, a future epidemic of a deadly disease.
So I have chosen to relegate the mathematical and computational details to appendices 
(which are meant to be accessible to anyone with some background in math modeling or computer science, with no prerequisites.)
I also eschew the usual Introduction/Methods/Results/Discussion format and adopt the form of an essay,
because I believe the flow of the argument will be easier for the general reader to follow.  

To begin, recall SARS, 
the acronym for ``Severe Acute Respiratory Syndrome,'' an atypical form of pneumonia 
caused by a coronavirus that generated a gobal pandemic in 2003.
The large-scale spread of SARS began on February 21 of that year, when a professor of medicine in Guangdong Province, China, who had been treating patients with pneumonia,
traveled to Hong Kong and checked into the Metropole Hotel downtown. For reasons still unclear,  
the Professor passed the virus on to about a dozen persons staying at the hotel.\footnote{There is a theory that he vomited on the rug outside his room, 
then tried unsuccessfully to clean it up---based on recovery of viral antigens from the rug, but not the room.} 
Then the Professor, feeling unwell, checked himself into the teaching hospital
and told the staff that he had a dangerous infectious disease and should be isolated, which the staff accomplished. (The Professor died of the disease in hospital.) 
When another patient from the 
hotel cluster arrived at the hospital, the staff made a medical mistake (administering an expectorant or performing aspiration), and subsequently every
doctor, nurse, medical student, and orderly who entered the room---$50$ or more---developed SARS. 
This was the first ``superspreader'' (SS) event of the epidemic. 
More such events followed: 15+ infected on a plane to Beijing, 37 at a hospital in Vietnam, and so forth. 
But the godzilla of all known SS events occurred at an apartment complex in Hong Kong called the Amoy Gardens, where a sick person visited for one night and 
generated 300+ secondary cases.\footnote{The 
individual apparently had diarrhea; the SARS 
virus, or perhaps a substrain, had a gut-tropism as well as lung-tropism.
Epidemiologists later explained the extraordinary transmission at the complex as the result of a design error in the plumbing system.
The bathrooms had floor drains that were supposed to be isolated by U-traps, but these only function if filled with water; but no one had told the residents to regularly dump
a gallon of water on their bathroom floors.}

Virologists eventually traced the source of SARS to bats, with an intermediate host, the civet cat, which was sold in live-animal markets in Guangdong Province.
The final toll (November 2002 to July 11, 2003, the date of the last known case) was: 29 countries, including Canada and the United States, affected;
8098 cases; and 774 deaths. 
(Thus the case mortality for SARS was around 10 percent; compare to the two percent estimated for the 1918 flu.)
The pandemic is thought to have been brought under control by introduction of better hospital procedures for rapidly isolating cases;
most clusters were ``nosocomial'' (occurring in, and specific to, a medical setting) with few secondary clusters, beyond those mentioned above, occurring in the community.

As I write, a new disease caused by a different coronovirus, called the Middle East Respiratory Syndrome (MERS) virus, 
has appeared and generated cases in Saudi Arabia, Jordan, Qatar, and other countries of the Arabian Penninsula 
(at this time, all cases outside the region have occurred in persons who had recently visited the Middle East).
A steady drumbeat of sporadic cases (91 since September 2012, as of 25 July 2013), 
some generating small SS events (mostly nosocomial), have been reported to the World Health Agency (and can be followed on the 
listservice of the International Society for Infectious Diseases called ProMed-mail). The current estimate of MERS case mortality is a frightening 50 percent (but ``case mortality''
has a denominator problem: the ``cases'' are hospitalized patients, and we do not as yet know how many milder cases are occurring in the community). 
 SARS, MERS, and (in one theory) HIV\footnote{Studies of transmission of HIV between an infected and an uninfected partner in a long-term relationship found very low transmission rates;
e.g., one transmission in 1,000 unprotected sex acts. This gave rise to the theory that HIV is an SS epidemic; the candidates for the superpreaders are: (a) persons in the primary
retroviral-infection period that lasts a few weeks, who have a thousand times the level of virus in blood and semen found in  chronically-infected patients; and (b) cases like ``patient zero,''
the Canadian airline attendant with an impressive Rolodex of sexual partners in many cities, 
described in `Randy Shilts's 1987 book, {\em And the Band Played On}.} all represent a different kind of epidemic
than usually treated in textbooks (particularly those that emphasize modeling). I will refer to a ``SuperSpreader epidemic,'' SSe for shorthand. 
Much of the conventional wisdom about epidemics does not apply to the SSe case.

Next, I informally describe models for an SSe, leaving details about software implementation and mathematical issues to appendices.
The appropriate kind of model is called a ``stochastic multi-type \bp.'' 
The adjective ``stochastic'' refers to random events, as in a dice game; in computer terms, when simulating the model the program makes 
calls on the \rng, abbreviated RNG (supplied with your operating system), when making updates. 
The ``\bp'' was introduced by mathematicians in the 1920s to describe, among other things, epidemics. (Other applications include demographic \pop\ growth and nuclear chain-reactions.)
In a \bp, some entity produces a (random) number of ``offspring'' at some rate, 
not depending on the number of other entities existing, for some (also possibly random) reproductive period; the offspring, which
may be of various types, can generate descendants in a similar manner. 
(The particular sort of \bps\ I use may differ from the kind described in math texts,
however; see the appendices for the details.) 
Mathematical biologists also introduced a deterministic model of epidemics dubbed ``SIR,'' for susceptible-infected-recovered, appropriate to describe a measles or influenza
epidemic. Here I am interested in interventions that bring the disease to a halt before it establishes an epidemic, 
so the susceptible \pop\ can be regarded as fixed. (By the time a moderate fraction of the
\pop\ of a major city or country had SARS or MERS, the pandemic would long since have been declared by WHO 
and, given the modern phenomenon of ``jet spread,'' airport closures and international panic would soon follow.) 
In addition, as the reader will understand from the results described below,
a deterministic model of an SSe is totally inappropriate.\footnote{Full disclosure: In  
March of 2003, the author and two colleagues sent a paper to
Science  proposing a stochastic \bp\ model of the SARS epidemic, 
concluding that the epidemic could be controlled by rapid isolation of infected patients if $\ro$ was not too large. 
Science rejected the paper, with the sole reviewer's comment being: 
``The model looks right, the conclusion looks right, but it doesn't penetrate!'' In fact, the outcome of the epidemic was as we predicted.
Later that Spring, Science published papers from two modeling groups; 
one used ODEs,~\cite{SARSdet}, which is absurd, and another,~\cite{SARSsto}, a model like the authors', with similar conclusions. 
The latter group had access to case incidence data from Hong Kong,
which our group lacked (we derived our model from published accounts, mostly in newspapers and on WHO and CDC websites). Later, the present 
author fit a stochastic model to the
Hong Kong data, see \cite{book08}, Chapter 6, in order to derive disease transmission parameters and demonstrate a new fitting technique.}
This is true for much of biology, despite the widespread use of deterministic equations, called by the ancient acronym ``ODEs'' 
whose significance nobody recalls, by mathematical biologists; see my book, \cite{book08}.  

Now we come to the famous ``$\ro$,'' also called the basic, or mean, reproductive number of the virus, which was historically the original ``tipping-point.''
It is defined as the average number of secondary cases caused (infected directly by) a primary case, in the absence of any treatment or intervention.  As we will see,
that number of secondary infections should be thought of as a \rv, so it has a variance as well as a mean; to uphold tradition, I denote it by $\vo$. 
Now we must carefully distinguish two epidemic scenarios, which I will call the ``uniform'' (or Poisson) and the SuperSpreader (SS) epidemics. In the former,
the number of secondary cases of a primary case (assumed for this discussion to have a fixed infectious period) is a Poisson \rv,
which is the name given in \prob\ theory for the number of events in a random but constant-rate accrual process, like the number of hits inside the ring by a darts-player
of little skill. The Poisson \rv\ is characterized by $\vo = \ro$ and a (super-exponential) 
fall off beyond the mean; e.g., if $\ro$ = 2, the probability of the primary case producing 10 secondary infections is
infinitesimal. Models in which such accruals have larger variances are said to possess ``\epv'' (EPV). 
What could be the meaning of this EPV? The more-infectious case---I will also refer to this person as a superspreader---might
have a special biological ability to spread the infection (perhaps through vomiting, diarrhea, or just a deep cough),
or might be situated in some place that facilitates transmission (e.g., in a crowded ICU, but not in isolation; on a plane; or visiting the Amoy Gardens). 

The simplest example, that I use throughout this paper for illustrations, is an epidemic model with two types of infected persons, each making a Poisson-distributed number
of secondary cases. The person of type one has an average number of secondary cases of $\rl$ (the subscript meaning lower-value), and the type-two person, $\rh$ (high-level);
these occur with probabilities $\pl$ and $\ph$, so the \pop\ average is:

\be
\ro \= \pl\,\rl \+ \ph\,\rh.
\ee

For instance, suppose $\rl = 0.7$ and $\rh = 30.0$, with $\pl =  0.9556$ and $\ph = 0.04437 $; then $\ro = 2.0$,
while (easy computation) $\vo =38.4$.

Mathematicians proved in the 1920s that a \bp\ with $\ro > 1$ can grow, while one with $\ro \leq 1$ eventually dies out. But it is often overlooked that a \bp\
with $\ro > 1$ {\em can nevertheless go extinct}. 
The outcome of a \bp\ with $\ro > 1$ (mathematicians use the phrase, ``supercritical'') is dichotomous: either heading for infinity or
destined to die out, with certain probabilities.
For instance, the process with the R's above plus assumptions about incubation periods, infectious periods, etc., resulting in various models;
(see the appendices) will disappear eventually with probability $\pe$ in the range $0.88$--$0.91$. For the uniform case, $\rl = \rh = 2.0$, $\pe$ is about $0.2$. 
As a rule, $\pe$ increases with increasing $\vo$ for fixed $\ro$.
This is easy to understand: the most probable ``index'' (epidemiology jargon for initial) case is a low-infectivity patient, and, although one or more secondary cases
might follow while the patient remains infectious (even if $\rl < 1$), the infection-chain is likely to die out unless preserved by the appearance of a superspreader.

See Table 1 for some illustrative examples. The simulation technique I used is described in appendix one, and the parameters are listed there in Table A. 
I also included results in the table from an exactly-solvable model, meaning explicit formulas exist for extinction probabilities (see appendix two), called the Markov case.
I included it to satisfy mathematicians and to check the software, by comparing \probs\ from repeated simulations to exact answers; 
but it has a feature that renders it dubious for use in biology. 
The Markov case is the model in which all waiting times---times to end of the incubation period; to end of the infectious period; and to generate the next secondary case---have
exponential distributions (see appendix two for the explanation of why the famous ``Markov property'' requires exponential distributions for all waiting times). 
But only for the last mentioned is this a realistic choice: the exponential law was derived from physics, where it represents radioactive decay; but people are not atoms. 
For applications I assumed, pending more data, normal distributions for the incubation and infectious periods (conditional on non-negativity, of course).
For these models, simulation must be used to compute \probs\ (for reasons spelled out in appendix two).
So the last column in the Table represents the biologically more-realistic case. 
The reason that the entries in this column are the smallest in the rows is that the Markov case has more EPV; see appendix two.
\centerline{}
\centerline{}
\centerline{}
\centerline{Table 1. Extinction \probs\ for some models and methods$^{*}$}
\begin{center}
\begin{tabular}{|c|c|c|c|c|}\hline
$\ro$ & $\rh$ & $\pe$ (Markov;formula) & $\pe$ (Markov;sim) & $\pe$ (normal;sim) \\ \hline
2.0 & 7.0 & 0.751 & 0.754  & 0.624  \\ \hline
 & 10.0 & 0.803  & 0.806  & 0.719 \\ \hline
& 15.0 & 0.851  & 0.857 & 0.788  \\ \hline
& 30.0 & 0.913  & 0.912 & 0.882 \\ \hline
1.2 & 7.0 & 0.936  & 0.933 & 0.889 \\ \hline
1.2 & 30.0 & 0.980  & 0.970 & 0.967  \\ \hline
\end{tabular}
\end{center}
$^{*}$ $\rl = 0.7$; normal standard deviations = 0.2 times means; 10,000 repetitions in simulations
\centerline{}
\centerline{}
\centerline{}

Next consider MERS,
for which (as I write) another sporadic case apparently shows up on average every three days. (A ``case'' means laboratory-confirmed in hospital, 
which may represent the very sick and potential superspreaders; there are probably more
cases appearing in the community.) 
Let ``Sp-int'' stand for the average interval between appearances of sporadic infections. Sp-int is hard to estimate, because it is difficult to distinguish 
sporadic from secondary cases; e.g., if in a family a father and two children become infected and the father becomes ill first, should the children be counted as secondary cases,
or were they exposed to the same external source of virus and so additional sporadic cases? Thus Sp-int might be considerably 
larger than three days. 
The law for first-time-to-a-Bernoulli-event gives:

\be
\hbox{average time of initiation of the pandemic} \= {\hbox{Sp-int}\over 1 - \pe},
\ee

\ni E.g., if $\pe = 0.98$ and Sp-int is 3 days, the mean initiation time is 150 days (but it could be much longer).
 
\begin{figure}
\rotatebox{0}{\resizebox{5in}{5in}{\includegraphics{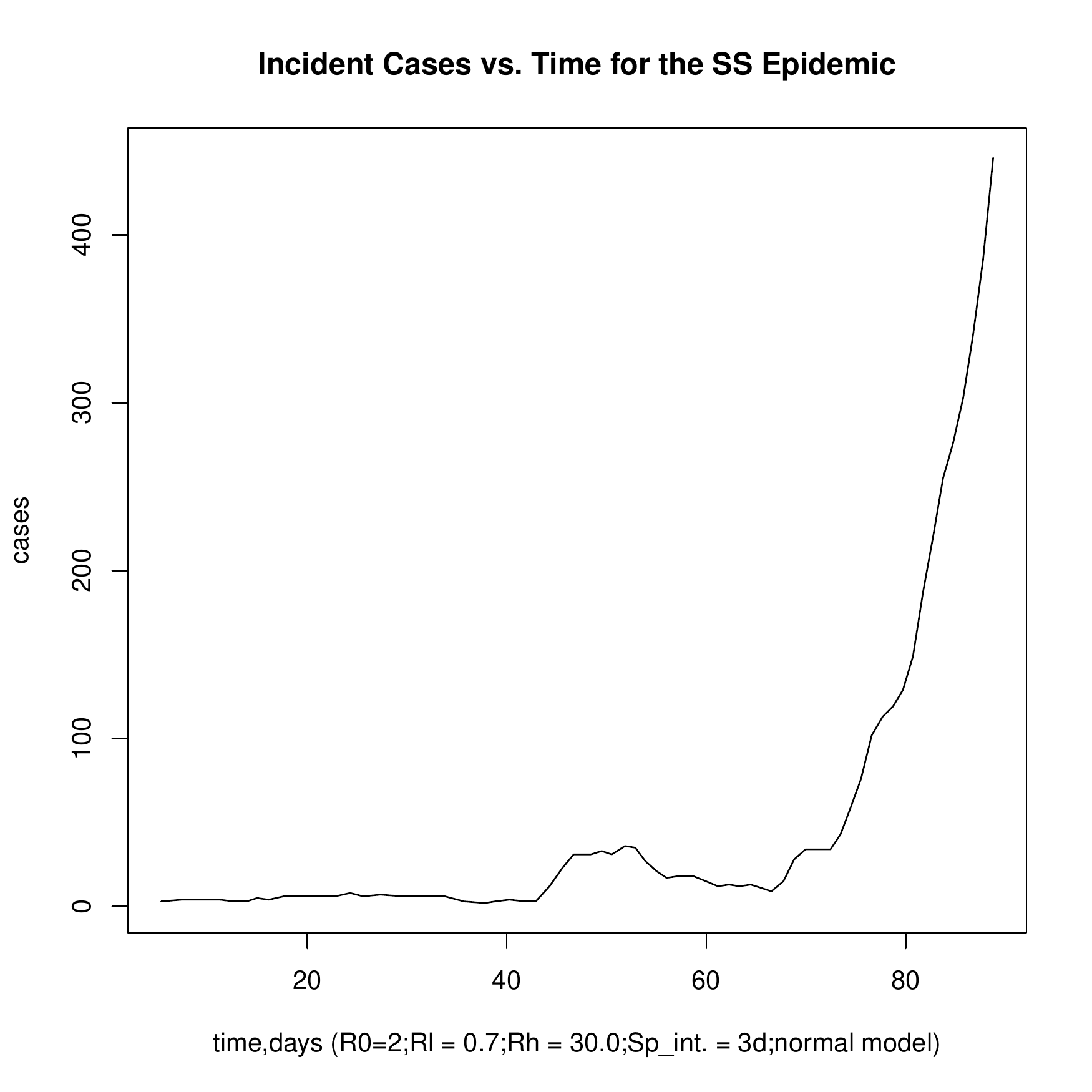}}}
\caption{An SSe with large $\vo$ and sporadic cases.\label{Figcase1}}
\end{figure}

\begin{figure}
\rotatebox{0}{\resizebox{5in}{5in}{\includegraphics{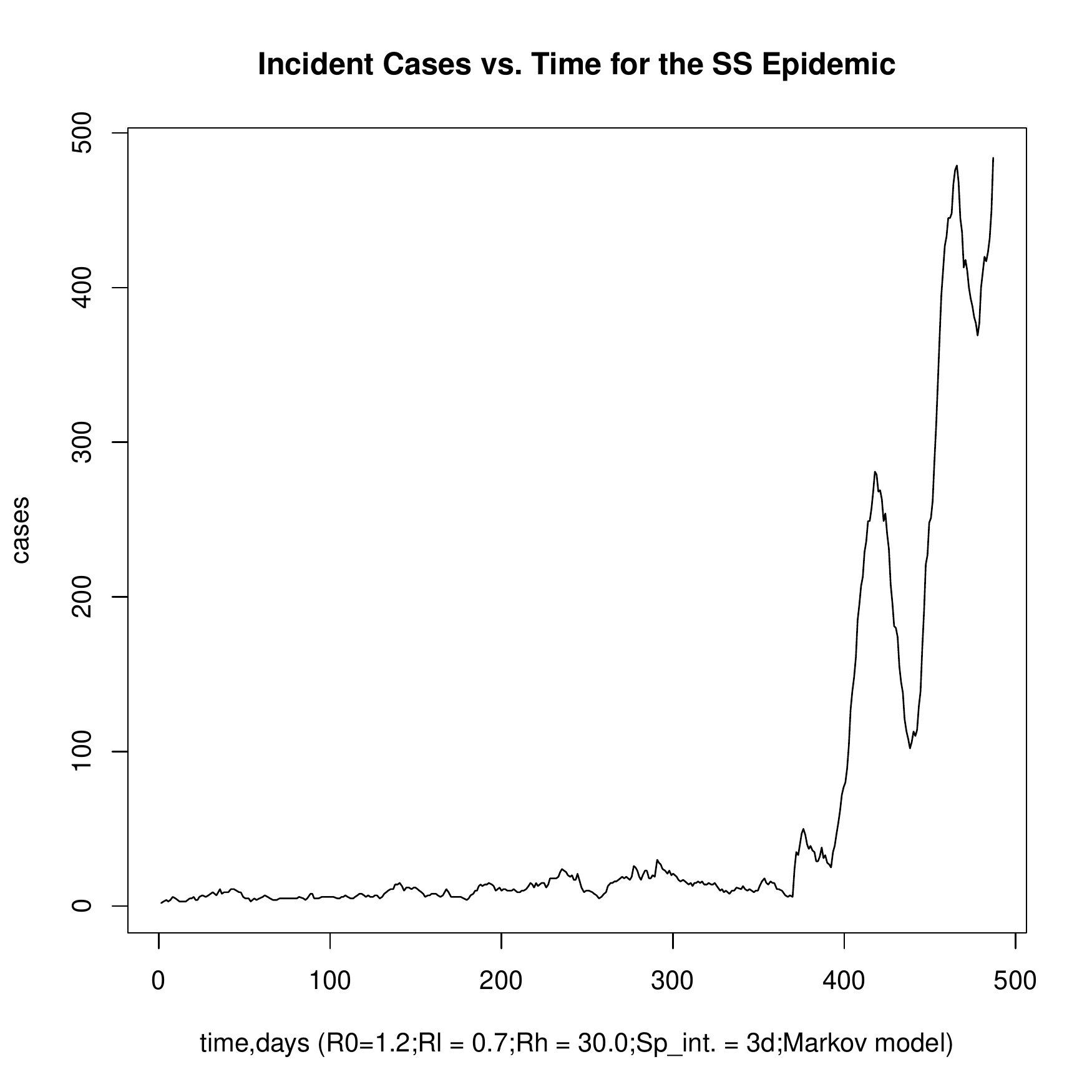}}}
\caption{An SSe with large $\vo$ but lower $\ro$, and sporadic cases.\label{Figcase2}}
\end{figure}

Because of the dichotomy and the tendency of most infection-chains to go extinct, the SSe with large EPV seeded by sporadic cases can be very surprising.
For instance, Figure \ref{Figcase1} shows an epidemic with R's as in the example above; 
for several months the epidemic seems indolent, then around 70 days, it takes off.
Figure \ref{Figcase2} shows another epidemic with parameters from the last row (most extreme case) of the Table; nothing exciting happens for a year,
which might cause premature optimism among public-health officials; then comes disaster.
(Rerunning with the same parameters but different ``seeds'' for the RNG, the timing of the pandemic in the figures is highly variable.)
Note that no spontaneous modifications, genetic or otherwise, are assumed in the virus for these simulations; disease parameters are fixed. 
The patterns are entirely due to EPV and chance.


Now to interventions. Neither vaccines nor therapeutic drugs currently exist for human coronaviruses; considering that the SARS virus has disappeared while MERS remains contained,
and the enormous cost of developing, testing, and licensing a new vaccine or drug, this unfortunate situation is likely to persist. Therefore, let us explore 
other interventions that could prevent an SSe pandemic by these viruses. The interventions are restricted to isolation of cases, preventing some secondary infections. The intervention might
be primary: rapidly isolating an infected patient in hospital in a negative-pressure room and requiring all attendant personel to wear gowns and P99 masks, hopefully decreasing
the patient's infectious period; or secondary:
tracing contacts of the primary case and getting as many as possible to go into isolation at home, thus diminishing secondary infections. The secondary intervention 
has one advantage over the primary: it can be instigated even if laboratory tests are delayed beyond the lifetime, or infectious period, of the patient, which may be a consideration
early in the epidemic. Both interventions have the same effect:
decreasing the effective ``R'' of the patient. An interesting special case, which I will refer to as the ``SS intervention'' (SSi), limits the intervention to high-infectivity cases.
Possible scenarios where the SSi is effective might include: SS patients are sicker and usually hospitalized, while low-infectivity cases remain in the community;
and where SS patients can be detected by some physiological or virological measurement. 

\def\IE{\hbox{IE}}
In most discussions of interventions in infectious-disease epidemiology, it is assumed that an efficacious intervention is one that drives $\ro$ below one. I argue that, for the SSe,
the proper quantity to look at is not $\ro$ but the {\em extinction \prob\ of the infection-chain}, which I have denoted by $\pe$. 
Thus the goal of the intervention must be to increase $\pe$;
indeed, I define \ie\ (IE) by:

\be
\hbox{IE} \= {\pei - \pe\over 1 - \pe}.
\ee

\ni where $\pei$ stands for the extinction \prob\ in the presence of the intervention. Thus $\IE = 1.0$ if the intervention drives $\pei$ all the way to one, and zero 
if $\pei = \pe$. Of course, if the \inter\ should push $\ro$ below one, then IE will be one, but IE could be, e.g., 0.75 even with an $\ro$ 
still greater than one after the \inter\ is active (as we will see below). IE depends on both $\ro$ and $\vo$, as we will see.

If the reader is dubious about this redefinition of efficacy, consider (a) the disease introduced into 1000 cities. If, e.g., $\pe = 0.7$ while $\pei = 0.9$ (efficacy, 67 percent; 
these numbers are not implausible, see Table 1), the effect of the
intervention is to spare 200 cities on average from a local epidemic. 
More convincing, perhaps, is to think about (b): the time before the epidemic takes off 
in a given region, for the 
sporadically-appearing disease. 
In the scenario in Table 1 fourth row, the epidemic takes off in a few months, 
but with the intervention increasing $\pe$ to the value in the last row,  
it takes roughly a half-year---perhaps  
enough time for epidemiologists or veterinarians to find and eliminate the source of the virus.

\begin{figure}
\rotatebox{0}{\resizebox{5in}{5in}{\includegraphics{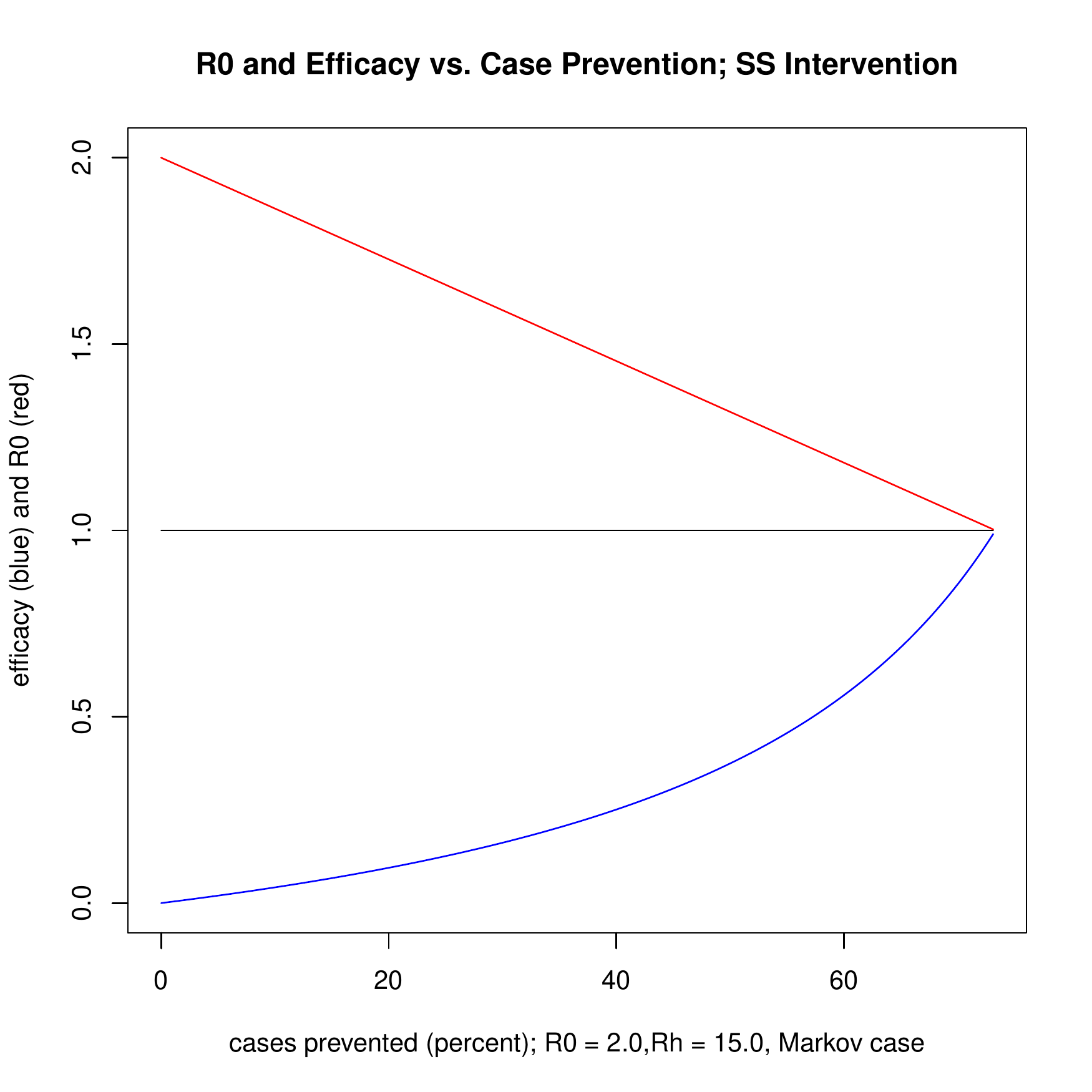}}}
\caption{Efficacy and effective $\ro$ under the SS intervention (Markov case).\label{Fig2}}
\end{figure}

\begin{figure}
\rotatebox{0}{\resizebox{5in}{5in}{\includegraphics{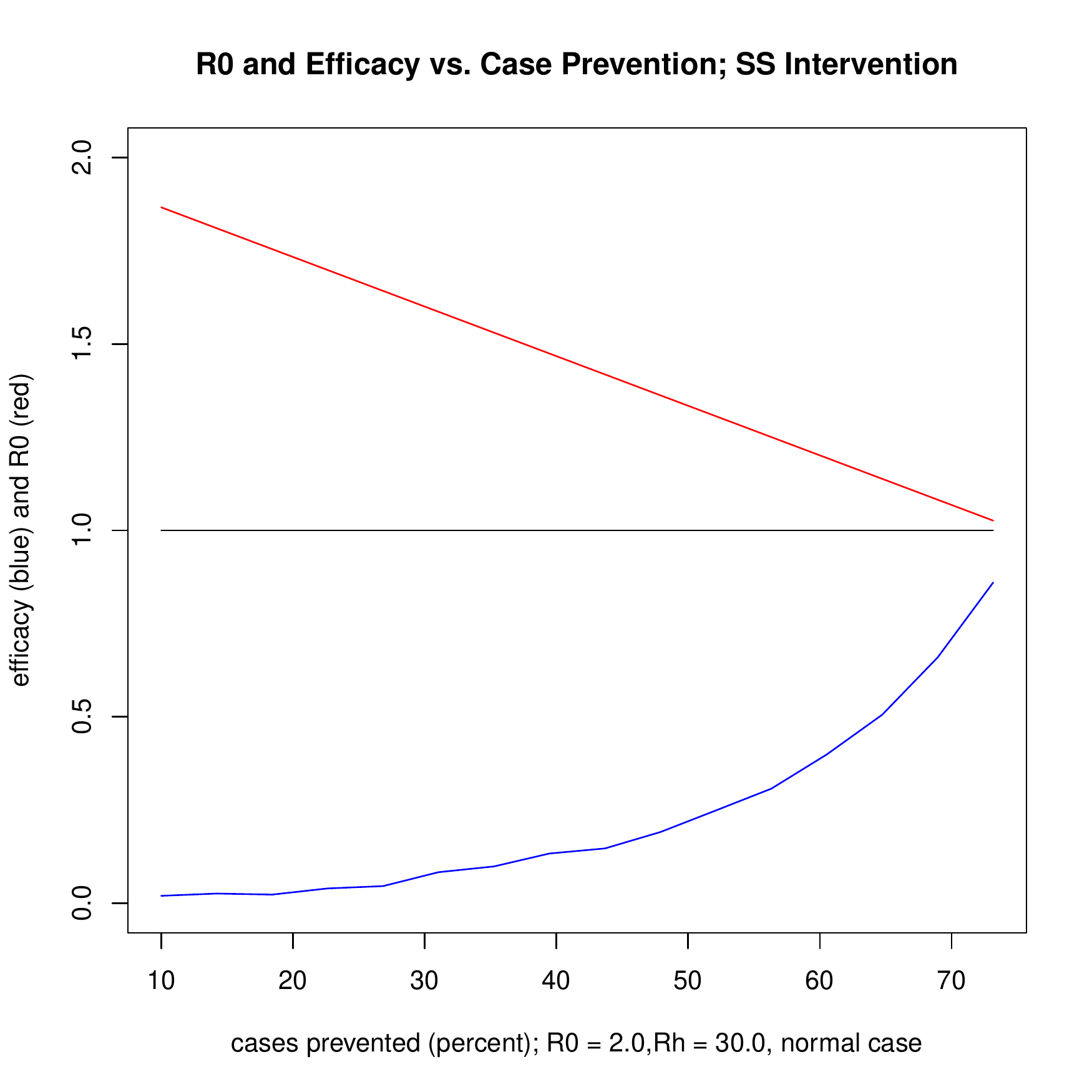}}}
\caption{Efficacy and effective $\ro$ under the SS intervention (normal case).\label{Fig3}}
\end{figure}

I therefore explored efficacy for such interventions through simulating an SSe epidemic with SSi's of various efficiencies (at lowering $\rh$). See Figures \ref{Fig2} and \ref{Fig3}. 
Efficacy is partially mediated by decreasing the effective (with-intervention) $\ro$, which in any case remains above one in the figures.
Note that, for a high-$\vo$ epidemic, even an intervention with efficiency in the 60-70 percent range 
can be quite valuable.

\begin{figure}
\rotatebox{0}{\resizebox{5in}{5in}{\includegraphics{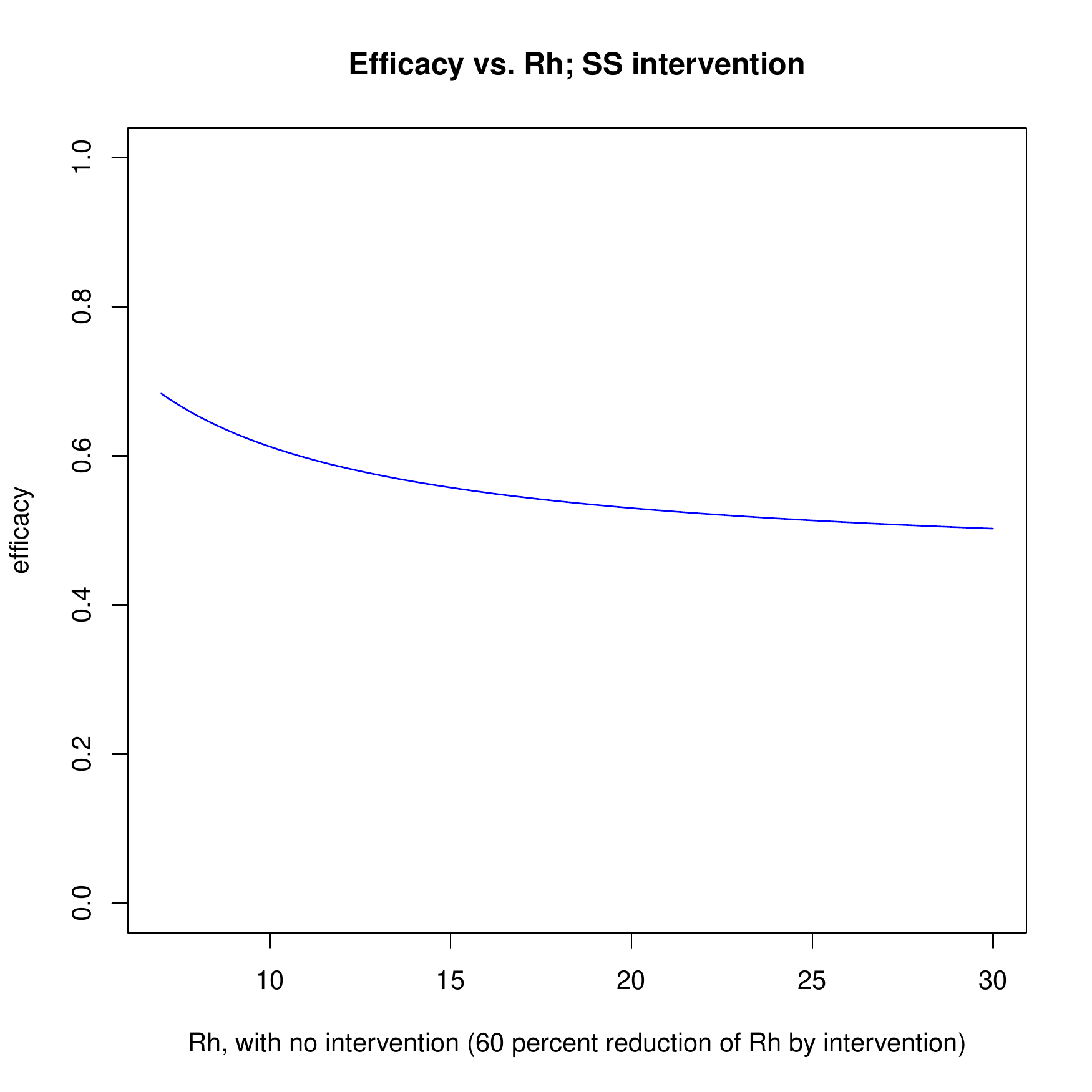}}}
\caption{Efficacy {\em vs}. $\rh$, with fixed SS intervention (Markov case; by formulas, see appendix 2).\label{eff_fig1}}
\end{figure}

\begin{figure}
\rotatebox{0}{\resizebox{5in}{5in}{\includegraphics{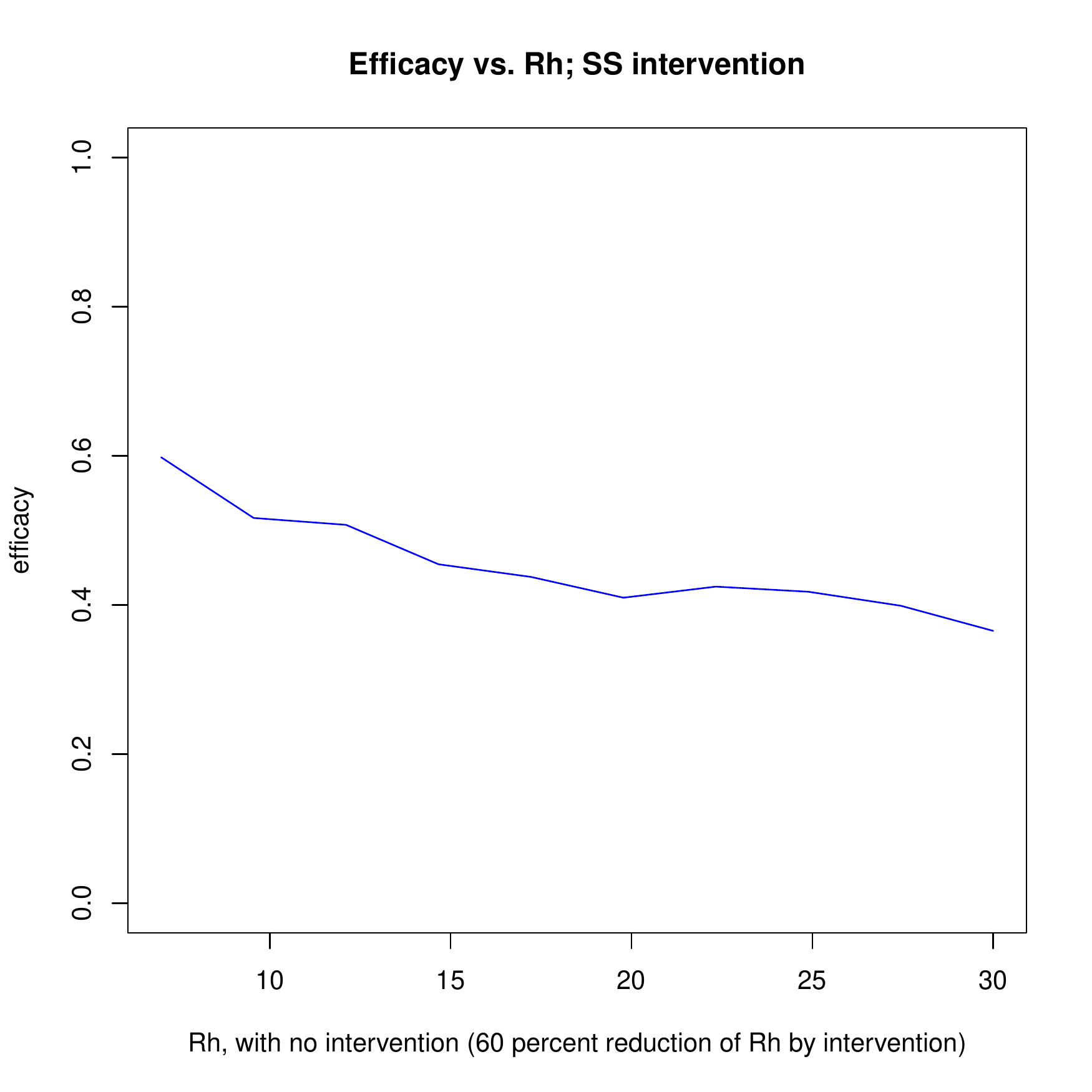}}}
\caption{Efficacy {\em vs}. $\rh$, with fixed SS intervention (normal case; by simulation).\label{eff_fig2}}
\end{figure}

I next asked whether it is easier or harder to stop an epidemic with a large $\vo$ (equivalently here, large $\rh$) than an epidemic with a smaller value, at fixed intervention.
The question is hard to answer through introspection or analysis. On the one hand, a larger $\rh$ should mean that the SSi intervention is more relevant; on the other hand, 
the explosive nature of case growth suggests the intervention would have to be larger. One thought suggests that efficacy should be an increasing function of $\rh$,
the other that it should be decreasing. 
See Figures \ref{eff_fig1} and \ref{eff_fig2}: in fact, efficacy in the models does diminish with increasing $\rh$ at fixed intervention efficiency, but falls off remarkably slowly.


As of August 2013, MERS has not generated a pandemic after 10 months of sporadic cases.
The obvious explanation is that $\ro$ is less than one---but
that could be an illusion; recall Figure \ref{Figcase2}! 
In July 2013, a group of epidemiologists published in Lancet, \cite{lancet} an estimate, based on 55 cases (there were more as of July but the authors culled some due to ambiguities), 
of $\ro$ as about 0.7, which I used for the lower figure ($\rl$) in the simulations. 
However, they made an ``Occam's Razor'' simplifying assumption 
which may have biased their analysis. 
Despite being critical of the authors' methodology, I agree with their conclusion that $\ro$ of MERS is currently less than one. 
(The best fit in my model-class to the cluster-data reported in \cite{lancet} 
had $\ro = 0.813;$ $\rl = 0.402$; $\rh = 17.9$; normal periods; and Sp-int = 3 days. I.e., an SSe but not a pandemic. See appendix three.)
Whatever the truth may be, in two months time 
roughly two million persons 
are expected to visit Saudi Arabia (the largest yearly gathering of people on the globe) for the annual 
religious pilgrimage called the hajj. (1.7 million visited in 2012.) Saudi Arabia has a population of 26 million, so 
this represents less than an eight percent increase; however, these pilgrims will be congregating in a few sites where there may be increased opportunity
for viral transmission from infected cases, possibly locally raising $\ro$ and, more likely, $\vo$. Also at present the source of the sporadic cases is under investigation, with reports of
a related virus in a bat (not found in the Middle East, however), 
and antibodies to coronaviruses in camels (but not in Saudi Arabia, and without recovery of a virus; because of cross-reactivity the latter report may be noise). 
The situation is too uncertain to make definite predictions about a MERS epidemic at this time.

Here is a summary of my conclusions about the SS epidemic and interventions, based on the modeling exercise. 
First, it is better not to be complacent about an often-fatal disease occurring at
low frequency and generating small clusters of secondary cases. Despite appearances, it may represent the ``kindling'' phase of a future pandemic---which may develop even in the absence of
genetic modification in the pathogen increasing transmission rates to or among humans
(as occurred a decade ago in SARS; but an event, given the current state of genetics, which nobody can predict). Second, the SSe has very different dynamics from more uniform 
epidemics caused by, e.g., influenza or measles; remarkably, this permits the design of interventions which are not available in the latter cases. For instance, because the
superspreader is rare and possibly detectable, a specifically targeted intervention may at least delay the onset of a pandemic long enough to allow epidemiologists to 
locate and eliminate the source of sporadic infections. Even detection after the end of a superspreader's infectious period can be efficacious (through contact-tracing).  
Finally, for my fellow math-modelers: the story related here is an instance of the important role played by heterogeneity, and stochasticity, in biology. (For other examples,
see my books \cite{book08} and \cite{book07}.) A desire for simplicity in modeling is not an excuse for overlooking these possibilities.

\section*{Appendix 1: the model}

The model is of ``agent-based'' type, which means that the characteristics of each case are stored separately in computer memory.\footnote{The distinction is with ``compartmental models,''
in which all the cases with similar characteristics are lumped into ``compartments,'' 
the terminology apparently derived from ``tissue compartments.'' ``Agent-based'' is computer-science jargon.} 
The descriptors of each case are: infectious status (in incubation period, so not infectious; or infectious); infectious type; and some integers assigned at creation for 
identification purposes (including the index of the primary, if a secondary case; and an indentifier for cluster). 
In addition, for each case some
floating-point numbers are stored: when a case is created in the incubation phase, a predicted (clock) time for entering the infectious phase; once in the latter, 
the time of creation of a secondary case (updated when that event occurs); 
and the time of death or end of infectious period. Because of the agent-based modeling, these times can be generated with any desired 
distributions. For the time-of-next-secondary-case, an exponential \rv\ is appropriate (consistent with the assumption that the infections caused in a fixed period
have a Poisson distribution). But an exponential \rv\ is a poor choice for the other waiting times 
(time-to-infectivity and to death or end-of-infectious period), because the exponential law has no memory.
Conventional mathematical models assume that all waiting times are exponentials, because the Markov Property then holds, permitting mathematicians to write down and solve equations 
for quantities like extinction probabilities; see the math appendix for discussion. However, the convenience of mathematicians should not be a determining element in science.
I assumed that the waiting times other than for secondary infections had normal distributions (conditional on being non-negative). 

The simulation starts by filling in the characteristics and times for the index case (and, for the sporadic scenario, a conjectured time for the appearance of the next sporadic case). 
Then the smallest waiting time is found; the corresponding changes are made, including
creation of a new case and its characteristics and waiting times if the event is a sporadic- or secondary-infection. The clock time is moved forward to this event time. 
At subsequent iterations the program runs through all the stored waiting times looking for the next one; then performs those changes, etc. 
In other words, the process is simulated in the most
straightforward manner, with no mathematical approximations whatsoever except for the belief that the RNG provides truly random digits (about 
which John von Neumann made his famous joke: ``Anyone who believes a computer can generate a \rv\ is living in sin''), which is unlikely to cause trouble here. 
When computing extinction probabilities, the process was run until either extinction occurred, 500 incident (incubation or infectious) cases existed on a single day 
(after which I assumed extinction was unlikely), or the clock time
reached four years. 10,000 samples were used in simulations. Table 1 reports the comparisons of the solvable (Markov) case with simulations of same; 
rerunning some cases with 100,000 repetitions, even the third decimal came out right.

Table A reports the parameters used. When more data about, e.g., MERS, becomes available, it will be interesting to fit the model in order to estimate parameters; a convenient method was
introduced in my book \cite{book08}, Chapter 5. The program was written in the C programming language, runs on anything, and is available from the author by request. 
(But if the reader is interested in modeling the SSe, it is always better to write your own program then rely on somebody else's.)

\centerline{}
\centerline{}
\centerline{}
\centerline{Table A. parameters used in the simulation}
\begin{center}
\begin{tabular}{|c|c|c|}\hline
parameter & value \\ \hline
$\rl$ & $0.7$ \\ \hline
$\rh$ & 7.0--30.0 \\ \hline
$\ro$ & 1.2--2.0 \\ \hline
incubation period & 5.36 days${^1}$ \\ \hline
infectious period & 3.83 days ${^1}$\\ \hline
standard error of normal periods & 0.2 x mean  \\ \hline
Spinterval & 3 or 6 days \\ \hline
\end{tabular}
\end{center}
$^{1}$ Derived from modeling of SARS in Hong Kong, \cite{SARSsto}, but because of the complexity of their model and fitting issues, these are questionable. See the discussion in my book
\cite{book08}, Chapter 6.
\centerline{}
\centerline{}
\centerline{}

\section*{Appendix 2: For mathematicians only}

I am an ex-mathematical physicist. So why the resort to programming the laptop? My motivation is simple: I intend to
use the model to make predictions about MERS and other SS epidemics, once transmission parameters become available. 
In my philosophy, the proper goal of any scientist
is to make predictions for future experiments or observations, because that is ultimately the only way to be sure that you are not modeling moonshine.
So Marc Kac's fabled advice for mathematical physicists (``mutilate, mutilate until you can solve the equations'') 
cannot be accepted. I may need to employ all sorts of peculiar distributions taken from empirical data. 
However, it is useful to solve even an unrealistic scenario because it provides a formula to compare to the output of the software for that case, 
helpful to allay the suspicion
that the results in this paper are due to bugs or numerical issues. 

The Markov case is the model in which all waiting times---times to end of the incubation period; to end of the infectious period; and to generate the next secondary case---have
exponential distributions. As pointed out in the text, only for the last mentioned is this a realistic choice. But it is required by the Markov property (``the future and the past are
independent, given the present'') because only the exponential is memoryless:

\def\ex{\hbox{Ex}}
\def\Pl{\hbox{P}\left[\,}
\def\Plk{\hbox{P}_k\left[\,}
\def\Plj{\hbox{P}_j\left[\,}
\def\Pr{\,\right]}
\be
\Pl \tau > t + s\,|\, \tau > s \Pr = \Pl \tau > t\Pr,
\ee

\ni provided $\tau$ is an exponential \rv: $\Pl \tau > t\Pr = \exp(-\lambda\,t)$. 
With the exponential for incubation and infectious periods, the number of secondary cases of a primary case is actually geometric rather than Poisson, with a larger
variance (so this case has EPV even even without multiple types). It is, of course, the Markov property that makes it easy to derive equations for \probs\
associated to stochastic processes. 

Here is a simple (non-rigorous?) derivation of the extinction \probs\ for the Markov version of the SS epidemic without sporadic cases.  
Let $\ex$ stand for the event: extinction of the infection-chain; $P_k$ for \probs\ generated by the process begun by a type-$k$ case; and $r_k$ for the corresponding
extinction probability, so:

\bar
\no r_k &\=& \Pl \hbox{extinction \prob\ given one initial case, type $k$} \Pr\\
 &\=& \Plk \ex \Pr.
\ear

Then by conditioning on the first secondary case, if there is one: 

\bar
 r_k &\=& \Plk \ex \,|\,\hbox{no secondary cases}\Pr \,\times\,\Pl \hbox{no secondary cases} \Pr + \label{first} \\
 && \sum_j \,\int_0^{\infty}\, \Plk \ex \,|\, \hbox{first secondary case is of type $j$, at time $\tau$}\Pr \\
 && \times\, \left[\hbox{temporal density of secondary case}\right]\,p_j\,
  \times\,\Plj \ex \Pr\,d\tau \label{sec} \\
 &\=& {\rho\over\rho + \gamma_k} \+ \big(\,{\gamma_k\over\rho + \gamma_k}\,\big)\,r_k\,\sum_j\,p_j\,r_j.\label{eqns}
\ear

\ni where $\gamma_k$ is the rate of generating secondary cases by a primary case of type $k$ and $\rho$ is the reciprocal of the time-to-end of the infectious period. 
Thus we obtain a quadratic system, easily reduced to a single quadratic for the two-level case of the text (below). However, the reader may doubt whether this is rigorous.

There is another (long-winded but fully rigorous) approach for the Markov case, which is to note that it is identical to a {\em Markov birth-and-death compartmental jump process}.
Here we have to distinguish incubation and infectious stages for each type; call the respective compartments (\pops) $X_k$ and $Y_k$. The rates of jumps for this process are:

\centerline{}
\centerline{}
\def\lra{\longrightarrow}
\begin{tabular}{ccc}
 & & \\
jump & rate & interpretation \\
&& \\
$X_k \lra X_k + 1$ & $p_k\,\sum_j\,\gamma_j\,Y_j$ & secondary infection; \\ 
$X_k \lra X_k - 1$ and & & \\
 $Y_k \lra Y_k + 1$ & $\eta\, X_k$ & progression;\\
$Y_k \lra Y_k - 1$ & $\rho\,Y_k$ & end infectious period.
\end{tabular}
\centerline{}
\centerline{}

Next given numbers $0 < r_k < 1$ and $0 < s_k <1$ introduce the moment-generating function (MGF):

\be
\phi \= E\,\prod\,r_k^{X_k(t)}\,s_k^{Y_k(t)} \equiv E\,\xi,
\ee

\ni where $E$ denotes expectation. Differentiating using Kolmogorov's forward equation (just a consequence of jumps occurring at given rates):

\bar
\no {\partial \phi\over \partial t} &\=& \sum_k\,\big\{\, p_k\,\left(\,r_k - 1\,\right)\,\gamma_j\,E\,Y_j\xi \+ \\ 
\no && \eta\,\left(\,{s_k\over r_k} - 1\,\right) \, E\,X_k\,\xi \+ \\
&&  \rho \left(\, {1\over s_k} - 1\,\right)\,E\,Y_k\,\xi.\big\}.
\ear

Now make the substitutions: $r_k\,\partial/\partial r_k\, \lra X_k$ 
and $s_k\,\partial/\partial s_k\, \lra Y_k$ to obtain the PDE:

\bar
\no {\partial \phi\over \partial t} &\=& \sum_k\,\left\{\, p_k\,\left(\,r_k - 1\,\right)\,\sum_j\,\gamma_j\,s_j\,{\partial \phi\over s_j} \+ \right.\\ 
\no && \eta\,\left(\,s_k - r_k\,\right) \, {\partial \phi\over r_k} \+ \\
&& \left. \rho \left(\,1 - s_k\,\right)\,{\partial \phi \over s_k}\right\}.
\ear

Solve this system by the method of characteristics; that is, define functions of time $r_k(t)$ and $s_k(t)$ to satisfy the ODE system:

\bar 
\no {dr_k\over dt} &\=& \eta\,\left(\,s_k - r_k\,\right);\\
 {ds_k\over dt} &\=& \gamma_k\,\left\{\,\sum_j\,p_j\,\left(\,r_j - 1\,\right)\,\right\} \,s_k + \rho\,\left(\,1 - s_k\,\right).\label{odes}
\ear

\def\br{ {\bf r} }
\def\bs{ {\bf s} }

Now note that $\phi$ is a function of $\br = \{r_k\}$, $\bs = \{s_k\}$  and $t$ and from the above

\be
\phi(\br(t),\bs(t);0) \= \phi(\br(0),\bs(0);t).
\ee

\ni Also from its definition, if we take as initial conditions $X_k = 1$ and $Y_k = 0$

\be
\lim_{t\to\infty}\phi(t) \= \Plk\ex\Pr = r_k,
\ee

\ni so setting the right-hand sides of the ODEs, (\ref{odes}), equal to zero to find the fixed point gives the required extinction probabilities.
The resulting equations are the same as (\ref{eqns}). 

For the two-level case, these equations are, using $R_k = \gamma_k/\rho$:

\bar
\no \left(\rl\,\psi - 1\right)\,r_1 + 1 \= 0;\\
\no \left(\rh\,\psi - 1\right)\,r_2 + 1 \= 0;\\
 \psi \= \pl\,r_1 + \ph\,r_2 - 1,\label{nn}
\ear

\ni which after a little algebra yields the quadratic:

\be
\rh\,\rl\,\psi^2 \+ \left(\, \rl\,\rh - \rl - \rh\right)\,\psi + 1 - \ro = 0.
\ee

\ni This is solved for $\psi$ (taking the negative sign in the quadratic formula and rationalizing the numerator) by:

\be
\psi \= { 4\left(\, 1 - \ro\,\right)\,\rl\,\rh\over \rl + \rh - \rl\,\rh + \sqrt{\left(\, \rl + \rh - \rl\,\rh\,\right)^2 - 4\left(\,1 - \ro\,\right)\,\rl\,\rh} }
\ee

\ni Note that if the extinction probabilities $r_1$ and $r_2$ are less than one, $\psi$ is negative and {\em vice versa}; hence that is the case if and only if $\ro > 1$.
Given $\psi$, $r_1$ and  $r_2$ can then be found from (\ref{nn}); $p_e = r_1$, $r_2$, or $p_1\,r_1 + p_2\,r_2$, depending on how you choose the initial condition. (The third is the realistic case.)
We used these results to reproduce tables and figures from the text, comparing to output obtained by
implementing exponentials for all waiting times and repeating the simulations.

Why can't we use the informal argument for the general case (without requiring everything to be exponential)? Because there were several hidden assumptions
in deriving (\ref{eqns}). For one, in (\ref{first}) we used the exponential integral:

\be
\Pl \hbox{no secondary cases} \Pr \= \rho\,\int_0^{\infty} \, du \, e^{- \rho\,u}\,e^{ - \gamma_k\,u},
\ee

\ni which in a non-exponential case would have to be done numerically. 
Worse, in (\ref{sec}) there was a concealed use of the Markov property; I erased conditioning as follows:

\bar
\no && \Pl \ex \,|\, \hbox{first secondary case is of type $j$, at time $\tau$}\Pr \= \\
\no && \Pl \ex \,|\, \hbox{first secondary case is of type $j$} \Pr,
\ear

\ni reasoning that the time of creation of the first secondary infection was not informative about the time left in the infectious period, which invokes 
the memorylessness of the exponential law, or just the Markov property. In a non-Markov case, that time would be informative, so one would have to
evaluate that conditional \prob\ as stated somehow, and (\ref{eqns}) becomes a system of integral equations which would have to be solved on the computer by
discretizing the time, solving matrix equations, then letting the discretization interval shrink, controlling for numerical error, etc. 
Relative to simulating the process on a fast computer, it doesn't seem worth the effort.

\section*{Appendix 3: The Lancet transmission paper, \cite{lancet}}

The Lancet authors, Breban \ea,  based their analysis on certain case-clusters (epidemiologically-linked cases), which they summarized in a table
(the scenarios concern how the authors interpreted the clusters in terms of sporadic vs. secondary cases):

\centerline{}
\centerline{}
\centerline{}
\centerline{Table B. Distribution of cluster sizes}
\begin{center}
\begin{tabular}{|c|c|c|}\hline
size & scenario 1 & scenario 2\\ \hline
1 & 17  & 11\\  \hline
2 & 4 & 2\\ \hline
3 & 3 & 3\\ \hline
4 & 1 & 1\\ \hline
5 & 0 & 2\\ \hline
24 &1  & 1\\ \hline
\end{tabular}
\end{center}
Converting from numbers of clusters to frequencies:
\centerline{}
\centerline{}
\centerline{}
\centerline{}
\centerline{Table C. Frequencies for cluster sizes}
\begin{center}
\begin{tabular}{|c|c|c|}\hline
size & scenario 1 & scenario 2\\ \hline
1 & 0.65  & 0.55 \\  \hline
2 & 0.15 & 0.10 \\ \hline
3 & 0.11 & 0.15 \\ \hline
4 & 0.038  & 0.05 \\ \hline
5 &  0 & 0.1 \\ \hline
24 & 0.038 & 0.05 \\ \hline
\end{tabular}
\end{center}

In their mathematical appendix, the Lancet authors begin by citing a paper in PLoS by others \cite{ploscb} 
who showed how to use \bps\ and Bayesian methodology to estimate $\ro$ from cluster data, 
allowing for some extra variance in reproductive number,
namely by assuming a certain distribution (called the ``negative binomial'') that interpolates between 
Poisson and geometric. In my terminology, those two cases represent the uniform epidemic with either
fixed or exponential infectious period---so much smaller EPV than allowed in my two-level model (which in fact has three free parameters, see last paragraph). 
The Lancet authors apply Bayesian analysis to the PLoS authors' distribution
but reject it as unable to resolve both parameters ($\ro$ and a dispersion parameter affecting $\vo$) because the latter ``could be very large, even 1000.'' 
It is here that they abused poor 
Occam\footnote{Friar William of Occam {\em did not} say that ``simplicity if best.'' 
He actually said, in the context of proving God's existence, that ``entities should not be multiplied
without necessity.''}
 and retreated to an analysis
using Poisson and estimating a single parameter, $\ro$. 
Of course, with this assumption the big cluster in the last row must include tertiary, etc., cases, because a cluster of 24 secondary cases of one primary case,
with an $\ro$ of 0.7, is virtually impossible. 

Here are simulated cluster frequencies under some models described in the text (10,000 simulations, each run to 91 cases, per model; a ``cluster'' consists of the tree
generated by a sporadic primary case):
\def\hb{\hfill\break}
\centerline{}
\centerline{}
\centerline{Table D. Simulated frequencies for cluster sizes}
\begin{center}
\begin{tabular}{|c|c|c|c|c|c|}\hline
cluster size & model A$^1$  &model B$^2$ & model C$^3$ & model D$^4$ & model E$^5$ \\ \hline
1 & 0.59  & 0.61 & 0.52 & 0.53 & 0.52 \\  \hline
2 & 0.14 & 0.14 & 0.17 & 0.17 & 0.16 \\ \hline
3 & 0.07 & 0.069 & 0.09 & 0.09 & 0.082 \\ \hline
4 &  0.041  & 0.041 & 0.055 & 0.053 & 0.048 \\ \hline
5 & 0.028  & 0.028 & 0.036& 0.035 & 0.03\\ \hline
6-23 & 0.11  & 0.098 & 0.123 & 0.112 & 0.091 \\ \hline
24+ & 0.013  & 0.0073 & 0.0084 & 0.004 & 0.066  \\ \hline
\end{tabular}
\centerline{}
$^1$ $\ro = \rl = \rh = 0.7$; Markov; Sp-int = 6 days.\hb
$^2$ $\ro = \rl = \rh = 0.7$; Markov; Sp-int = 3 days.\hb
$^3$ $\ro = \rl = \rh = 0.7$; normal, st. error = 0.2 times mean; Sp-int = 6 days.\hb
$^4$ $\ro = \rl = \rh = 0.7$; normal, st. error = 0.2 times mean; Sp-int = 3 days.\hb
$^5$ $\ro = 1.2, \rl = 0.7, \rh = 30.0$; normal, st. error = 0.2 times mean; Sp-int = 6 days.\hb
\end{center}
\centerline{}
\centerline{}
 
Not surprisingly, the uniform (Poisson) model with $\ro = 0.7$ has difficulty fitting the big cluster of 24 cases in the last line of Tables B/C;
indeed, it appears as if Models B-D can be rejected, say by a Fisherian p-value test.
This suggests that what happened in the Lancet authors' Bayesian fitting procedure is that their single-type, Poisson likelihood settled for fitting the small clusters and ignored the the big one.
Which could render their logic circular ($\ro$ is less than one because, in a model with no higher-infectious types and ignoring one data point \dots $\ro$ comes out less than one).

However, does that mean the authors' conclusion is wrong? To answer that question, 
I utilized the NkNN method (advertised in my book \cite{book08}) to search for best-fitting models in my model-class. 
This venerable method selects among various theories, given samples from each, that best explains a given data-set; it is commonly used in pattern-recognition. 
(It reduces to maximum likelihood in the limit of infinitely-large samples from each theory; i.e., if you have an infinitely-fast computer to perform the simulations.)
Here the samples were 1,000 simulated epidemics, and corresponding cluster-frequencies, from models chosen at random with $0.2 < \rl < 1$, $1 < \rh < 30$, and $\rl < \ro < \hbox{min}(\rh,2.0)$,
with 1,000 repetitions. The data-set was column one of Table C.
The search turned up excellent fits for the
frequency data (smallest sum-of-squares, 0.052):

\centerline{}
\centerline{}
\centerline{Table E. Simulated frequencies for cluster sizes}
\begin{center}
\begin{tabular}{|c|c|c|}\hline
cluster size & model F$^1$  &model G$^2$ \\ \hline
1 & 0.67  & 0.680  \\  \hline
2 &0.16 & 0.17 \\ \hline
3 & 0.062  & 0.066 \\ \hline
4 & 0.027  & 0.031 \\ \hline
5 & 0.013 & 0.016 \\ \hline
6-23 & 0.030   & 0.022  \\ \hline
24+ & 0.032  & 0.013 \\ \hline
\end{tabular}
\centerline{}
$^1$ $\ro = 0.813;$ $\rl = 0.402$; $ \rh = 17.9$; normal; Sp-int = 3 days.\hb
$^2$ $\ro = 0.539$; $\rl = 0.392$; $ \rh = 20.7$; normal; Sp-int = 6 days.\hb
\end{center}
\centerline{}
\centerline{}

On the basis of this result, I must agree that $\ro < 1$ for MERS at present.
Note however that the good fit of Table E to Table C is attained because an SS model reproduces the big cluster in the last
row of the latter. 
The Lancet authors' choice to limit their model-class to a Poisson distribution for secondary cases ruled
out this possibility---a SuperSpreader, multiple-levels, epidemic model, but with $\ro$ less than one---{\em a priori}.
On the other hand, ``fitting'' a three-parameter model 
to data consisting of seven frequencies cannot be recommended either.

Now a bit of preaching about modeling traps. When you have a model with free parameters and a data-set, there are two ways to delude yourself.
Methods like Bayesian analysis, possible if you have an explicit likelihood, or my NkNN method, which works for anything provided you can simulate fast enough,
will generate a ``best-fitting model.'' If some parameter (like the PLoS authors' dispersion parameter) fluctuates so much that it can't be pinned down, 
you can choose to omit it---i.e., mutilate the model in order to be able to ``estimate the parameters.'' 
Or you can keep the troublesome parameter and let the software choose a ``best model.'' 
Either way, the model selected may be false; but you won't discover that until more data arrives.
In my book, \cite{book08}, I labeled these two pitfalls the ``Simple-is-Best Trap'' and the ``Etch-a-Sketch Trap''
and argued that the modeler is required to steer between these two hazards, like Odysseus sailing between Skylla and Charybdis. 
Fall in the first, and your parameter estimates will be wrong or biologically unintelligible; fall in the other, and your model can ``explain'' anything (which Karl Popper
decried as not science). 
Ultimately, the best approach is to make
a prediction---even when it fails, you will have learned something. (Unfortunately, on the basis of the data thus far available, making a prediction about MERS is impossible.)


\begin{thebibliography}{10}

\bibitem{lancet}
Breban, R, Riou, J, Fontanet, J. Interhuman transmissibility of Middle East respiratory syndrome coronavirus: estimation of pandemic risk.  
Lancet, published online July 5, 2013.

\bibitem{ploscb}
Blumberg, S and Lloyd-Smith, JO. Inference of $\ro$ and transmission heterogeneity from the size distribution of stuttering chains.
PLoS Comput Biol 2013: 9, e1002993. 

\bibitem{SARSdet}
Lipsitch M, Cohen T, Cooper B, Robins JM, Ma S, James L, Gopalakrishna G, Chew
SK, Tan CC, Samore MH, Fisman D, Murray M.
Transmission dynamics and control of severe acute respiratory syndrome.
Science. 2003 Jun 20;300(5627):1966-70. Epub 2003 May 23.

\bibitem{SARSsto}
Riley S, Fraser C, Donnelly CA, Ghani AC, Abu-Raddad LJ, Hedley AJ, Leung GM, Ho 
LM, Lam TH, Thach TQ, Chau P, Chan KP, Lo SV, Leung PY, Tsang T, Ho W, Lee KH,
Lau EM, Ferguson NM, Anderson RM.

Transmission dynamics of the etiological agent of SARS in Hong Kong: impact of
public health interventions.
Science. 2003 Jun 20;300(5627):1961-6. Epub 2003 May 23.

\bibitem{book08} 
Wick, WD. {\em Fitting Non-Linear, Stochastic Models to Data in Biology and Medicine}. Available on Amazon.com, 2012.

\bibitem{book07}
Wick, WD and Yang, OO. {\em War in the Body: the evolutionary arms race between HIV and the human immune system and the implications for vaccines}. 
Springer, 2013.

\end{thebibliography}
\end{document}